# SMART LAPTOP BAG WITH MACHINE LEARNING FOR ACTIVITY RECOGNITION

Dwij Sukeshkumar Sheth, Shantanu Singh, Prakhar S Mathur, Vydeki D

*School of Electronics Engineering (SENSE), Vellore Institute of Technology Chennai*

*Abstract*—In today's world of smart living, the smart laptop bag, presented in this paper, provides a better solution to keep track of our precious possessions and monitoring them in real time. As the world moves towards a much tech-savvy direction, the novel laptop bag discussed here facilitates the user to perform location tracking, ambience monitoring, user-state monitoring etc. in one device. The innovative design uses cloud computing and machine learning algorithms to monitor the health of the user and many parameters of the bag. The emergency alert system in this bag could be trained to send appropriate notifications to emergency contacts of the user, in case of abnormal health conditions or theft of the bag. The experimental smart laptop bag uses deep neural network, which was trained and tested over the various parameters from the bag and produces above 95% accurate results.

## I. INTRODUCTION

The number of smart devices is expected to be around 20 billion by 2020. With the rising number for smart devices and advancements in computing nowadays we can find many applications of Internet of Things (IoT) in the common devices and things that we use in our daily day to day life. Nowadays there are many devices that people use like smart watches, activity trackers etc. Smart laptop bag is one such solution that contributes to the field and is a potential device for the masses [11].

The smart laptop bag combines the features of IoT (Internet of Things) with the capabilities of advance and smart computing which helps to make the device intelligent like other smart devices. It is an all in one solution for most of the features that the user desires, like activity tracking, location tracking, safety feature etc. which have become a necessity for the modern era. The bag includes sensors for measuring the quantities like acceleration, yaw, pitch, roll, load on shoulder, temperature, humidity, presence of specific gases in very high quantity that may be injurious and may have gone undetected by the human nose but can cause side effect etc.

This paper is organized as follows. Section II presents the related systems in literature. In the Section III we discuss about the comparison of the proposed model with other related solutions proposed in the past, Section IV provides the computing techniques applied to this work. The hardware design of the system is described in Section V. Details about the software and the user interfaces for the same are given in section VI. The analysis part performed on the data using a machine learning approach is described elaborately in Section VII. The Conclusion and Future scope of work possible in the proposed model has been proposed in Section VIII.

## II. RELATED WORK

As inferred from the literature survey, recently there has been some work on Smart laptop bags as well in the previous years which have led to rise of some products in the market like bags that come with inbuilt trackers etc. Apart from that there are some prototypes of bags that include a safety button for security purpose which are yet to hit the market in due time. Usually these solutions just aim to target a specific issue for which it is designed. There seems to be a lack of a solution that can address all these issues as well as include some smart features that can provide intelligent analytics and services to the user [5 6].

## III. COMPARATIVE ANALYSIS

The research work on Multipurpose Smart Bag as performed by Shweta M (2016) [5] talks about the design on a smart bag that incorporates the use of several hardware devices like RFID, LCD, Bluetooth and SOS button. Shweta M (2016) concluded the design of the multipurpose bag with inclusion of a safety application which connects to the bag using an Bluetooth module and sends out SOS messages to emergency contacts and the buzzer mounted on the bag will start emitting sound signals. However, the model proposed does not incorporate the use of any other sensing device to sense other parameter in case of an emergency and provide advanced analytics so that additional help can be made available when needed. Moreover, the proposed model of the system suggested by Shweta M (2016) has made an assumption that in case of any kind of emergency the police must be contacted apart from the emergency contacts stored by the user.

Similarly, Athul P Anand (2016) proposed a similar model as Shweta M (2016) of a smart bag specifically designed for students where they try to provide the feature of a panic button for the student which can be pressed in case the student feels insecure so that the parents are notified about this state and the GPS coordinates of the student are sent to the parent [6]. However, this model lacks the use of predictive models which can be helpful in providing immediate help to the user in such case until the parents can actually reach the student out based on the data on which the predictive model has been trained.

Hence, our proposed model the smart laptop bag tries to fix the missing functionalities as spotted in the above described models so that we can provide the user with additional functionality using the advances in smart computing and incorporating intelligence in the model using a machine learning approach.

## IV. TECHNICAL BACKGROUND

The system uses the following computing methods and techniques to perform the task it is supposed to:

*A. Deep Neural networks*

A Deep Neural network is a Neural Network with several hidden layers in its model that add up complexity to the neural network. Any neural network with more than two layers is classified to be a Deep Neural network. These types of networks have an input layer, an output layer with at least one hidden layer in-between. Each layer performs specific types of ordering and sorting in a process that is sometimes referred to as "feature hierarchy." [8].

*B. Google Firebase*

Firebase is a service that enables users to develop Web/Mobile applications with minimal server-side scripting so that the development process is easier and faster. By using firebase service, we do not need to create REST API's for interaction with our frontend service, like in our case, the mobile application installed on the user's mobile device. Firebase supports Android, Web, iOS, OS X clients. Firebase helps us by storing the data from the sensors and implement the classification using the machine learning model that we have developed using the ML Toolkit feature [10 16].

## V. HARDWARE OF THE PROPOSED SYSTEM

*A. System Description*

The project aims at designing a Smart laptop bag that is capable of providing the user with advanced safety, protection and monitoring system. It helps the user to keep a track of the position and state of the bag by considering things like Geographical location, water seepage into the bag, load on the users shoulders due to the straps, activity tracking and classification into the specific type using Machine learning approach. Based on the values that are recorded the appropriate notifications can be sent to the client-side app which is supported by Google's Firebase – An Online Cloud service where the data is logged by the bag. Apart from monitoring there are several safety features that have been incorporated in the bag like a SOS button which can be triggered to get help in case of emergency. This can be done using the mobile application which will trigger an alarm that can help the user find the bag if they have misplaced it somewhere.

When the user wears the bag, the sensors are mounted in such a way that they can measure the required parameters optimally and send the data to the microcontroller unit which will be present inside the bag itself along with the power unit which provides power for the entire system to operate. The microcontroller unit is further connected to a Raspberry pi which helps in collecting the data from the bag and send it to the cloud. In addition to sending data to the cloud the Raspberry pi connects to the user's Mobile phone in order to trigger the updates and use the packet data to send the data over the cloud. Once the data is stored in the cloud it is retrieved by the client-side mobile application where we will have our machine learning model saved which will classify the user's activity and also display the relevant data to the user in the application which will be installed on his mobile device. The prototype of the same has been implemented. The use of machine learning, Android technology, combined with the recent advances in smart computing could be the key to solve emerging problems and provide a plethora of smart devices which can assist the users in several ways and ease the difficulties faced by one, on a regular basis.

Sensors including GPS module, Water Sensor, Temperature and Humidity Sensor, Piezo Sensors, Gas sensors, Accelerometer, sense the corresponding parameter. All these sensors are connected to Arduino Board, which collects all the parameters from sensors and sends it serially to Raspberry Pi module. Fig 1 describes the schematic diagram of the entire system for the case discussed in this section and was modelled using fritzing which is an open-source simulation tool [14]. The Raspberry pi creates a JSON format for all the parameters and pushes it to the Firebase (Database). Android app fetches the data from Firebase processes it and displays the parameters and notification in case of any event.

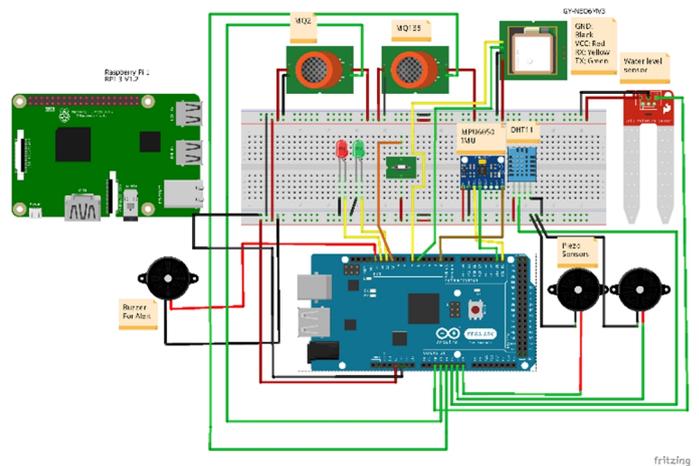

*Fig 1. Hardware schematic diagram for the system*

*B. System Impelementation*

This section describes the implementation along with the working of our system in a real-life scenario. The complete details about the various parameters that we are trying to collect as well as the analytics that we are performing on it to provide the optimum output has been described in this section. Fig 2 gives an overview of the components and the interactions among them while collecting the data. Fig.3 represents the way in which the data is organized and processed for providing the required output to the user.

*1) Block Diagram*

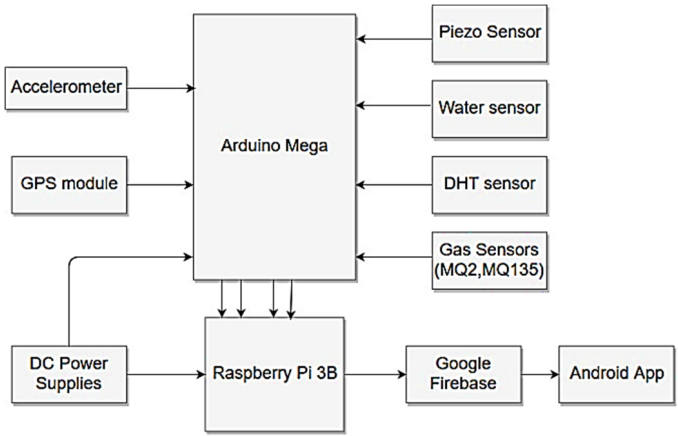

*Fig 2. A Block Diagram representing the components of the proposed system.*

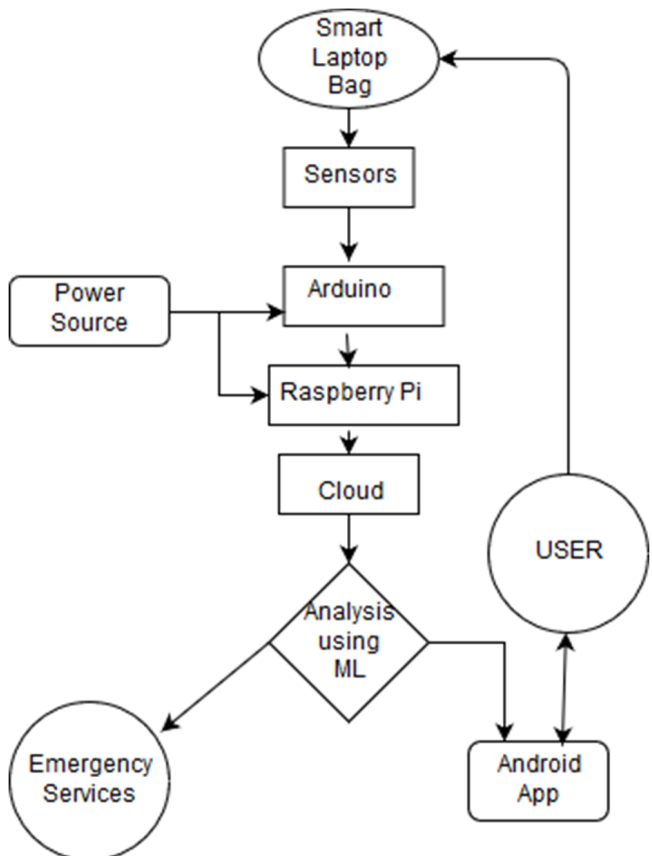

*Fig 3, Diagram representing the components of the proposed system for analysis and management of data.*

C. Microcontroller Unit

For our application we are using an Arduino Mega 2560 Microcontroller board. The Arduino Mega 2560 is a microcontroller board based on the ATmega2560. It consists of 54 digital input/output pins (15 of which can be used for PWM outputs), 16 analog inputs, 4 UARTs (hardware serial ports), a 16 MHz crystal oscillator, a USB connection, a power jack, an ICSP header, and a reset button [1].

D. Raspberry PI

The Raspberry Pi is a small, credit-card sized computer that runs on an ARM11 microcontroller @1GHz and consists 1GB of RAM memory. The board can be connected to a computer display or an TV using the HDMI interface port which is inbuilt on it and interface it with a standard keyboard and mouse for interaction with the user. The board is powered by a Linux based operating system, in our case we are using Raspbian OS. Linux is a great match for Raspberry Pi because it's free and open source [2]. It has the same functionalities as a desktop computer does at a reduced performance. It can support up to 32GB of external memory and consists of 4 USB Ports [3].

E. Sensors

*1) GPS Module (Neo-6M)*

GPS stands for Global Positioning System and it gives us the current date & time, latitude, longitude, speed, altitude and the headin direction. It can be interfaced with a normal microcontroller using a 5V GPIO interface which can be achieved directly in case of microcontrollers like Arduino or using an 3V-5V convertor unit in case of some mbed boards like the LPC11U24. This GPS module can act as an standalone component as well when provided with a 5V supply. The chip comes with an inbuilt RTC backup battery and can be directly connected to USART (Universal Synchronous Asynchronous Reciever Transmitter) of the microcontroller.

*2) MPU6050*

The InvenSense MPU-6050 sensor contains a MEMS accelerometer and a MEMS gyroscope on a single chip. It is accurate and contains 16-bits analog to digital conversion hardware for each channel. Therefore it captures the x, y, and z channel at the same time. The sensor uses the I2C-bus to interface with the Arduino. It is cheap considering the fact the is has both an accelerometer as well as a gyroscope on the same chip. The MPU6050 interfaces to the arduino microcontroller using an I2C BUS interface where the sensor acts as an slave whereas the Arduino Acts as then master device and retrieves the necessary data from the same. Using the data from the MPU6050 we can get to know the activity that the user is performing using the accelerometer and the gyroscope sensor values and classify it into classes using the machine learning model which is describe in the sections ahead.

*3) MQ2 Gas Sensor*

MQ-2 gas sensor is used to sense the presence of LPG, Propane, Hydrogen, Methane and other combustible gases that are said to be harmful and usually go un-detected. It is a low-cost sensor and is suitable for different application and industrial use. The sensitive material of MQ-2 gas sensor is $SnO_2$, $SnO_2$ has a lower conductivity in clean air. When the target combustible gas exist, the sensor's conductivity increases along with the increase in the gas concentration [4].

*4) MQ135 Gas Sensor*

The MQ-135 gas sensor is used for the detection of $NH_3$, $NO_x$, alcohol, benzene, smoke, $CO_2$, etc. It is an low-cost sensor that is an essential part of every air quality monitoring device like smoke detectors etc which are used in industries as well as places like airports, malls etc [4].

*5) DHT11 Sensor*

The DHT-11 sensor provides the ambient temperature and humidity values as read by it which will help keep a track of the same to monitor the environment in which the user is present. The sensor operates on 5V and can interface with the Arduino microcontroller using an analog General-Purpose Input-Output (GPIO) pin.

## VI. SOFTWARE OF THE PROPOSED SYSTEM

### A. FIREBASE BACKEND SERVICE

For sending the data to the firebase server we will be retrieving the data from the Arduino onto the raspberry pi where we will be running a python script that retrieves this data from the Arduino over UART communication and send it to the firebase cloud using PATCH requests. Once the data is pushed every 2 seconds the new values of the sensor are indexed immediately into the database and the machine learning model is used to classify it and send all the results to the android application [10].

### B. MOBILE APPLICATION

We are using a custom-built android mobile application that fetches the data from the firebase server and displays it to the user along with the activity details as classified by the machine learning model which will be discussed in the further sections of this article. The Mobile application is built using the Android Studio application.

The Application consists of a login screen, which is depicted in Fig. 4. The login interface is used for authentication where the user can enter his/her credentials in order to see the data which is being fetched by the application from the firebase backend service. Once the authentication stage is passed the user can view the data that is collected from the system and indexed in the storage of the Firebase service, a view of the same can be seen in Fig. 5.

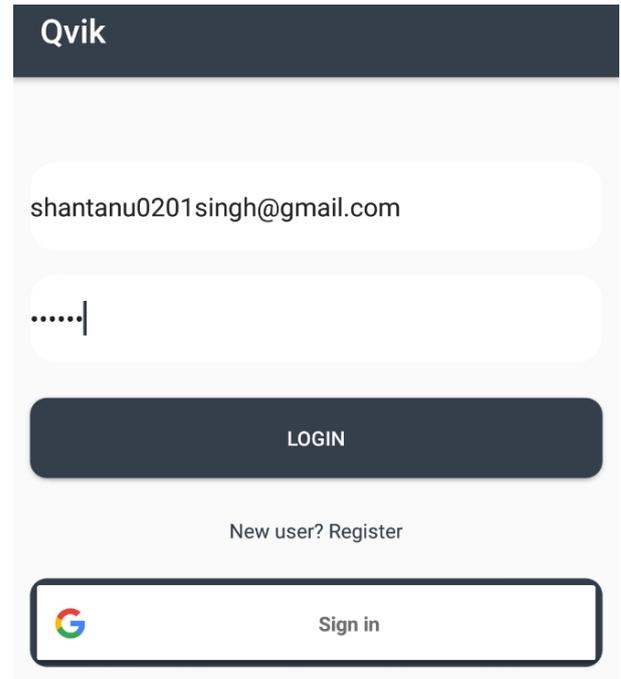

*Fig 4. A view of the login interface*

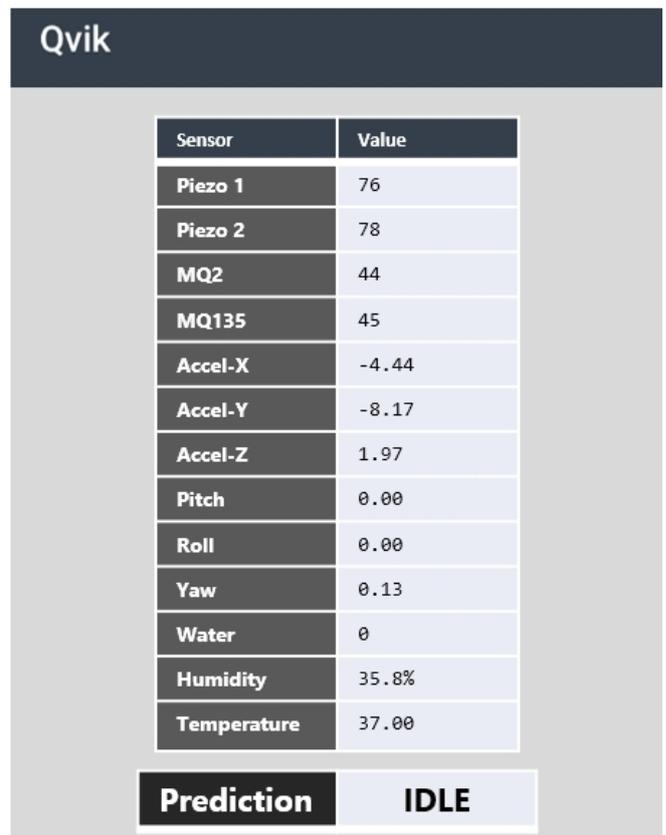

*Fig 5. A view of the android application*

## VII. MACHINE LEARNING

### A. Use of the Machine Learning Model

The use of the machine learning model here is to detect the type of activity that is performed by the user when he/she is wearing the bag. The machine learning model uses the sensor data which is retrieved from the sensors that are mounted in the bag and uses a deep neural network to classify the type of activity that is being performed. Based on the type of activity classified by the machine learning model the android app acts on the same and sends appropriate notifications to the user or his/her emergency contacts in case of an emergency. The same can be done by hard coding the thresholds for each sensor and programming the logic for the activity classification on the microcontroller but it may result in a limited number of output classes and many misclassifications as compared to the machine learning model hence to solve this problem and providing a better experience the machine learning model comes into this application and helps in performing the identified task.

### B. Dataset

The data used consists of 13 features with values from all the sensors used on the bag like the piezo sensors, gas sensors, accelerometer & gyroscope etc. Every data field is further labelled into the corresponding activity type. Some of the activity classes present in the dataset are: Walking, IDLE, Running etc. The data was collected from the system in real-time and labelled based on the activity and further stored in an csv file so that it can be used for training the model. The complete dataset consists of 1743 data entries which are labelled.

The details of the person who used the bag during data collection are:
- Weight: 66 Kilograms
- Height: 170.2 Centimetres
- BMI: 22.8

A random selection of 90% of the dataset was used as the training set while the remaining 10% was used to test the accuracy of the model.

### C. Training the Model

Here we have split-up the data into training and test set. For the model we are using a Deep neural network with 5 hidden layers having 15,20,25,30,60 neurons in each hidden layer respectively. Every hidden layer uses the ReLU (Rectified Linear Unit) as its activation function. The input layer consists of 10 neurons with ReLU as activation function [7]. The Output layer consists of 5 neurons and uses the SoftMax activation function. The model has been trained for 10 epochs with a batch size of 128 samples and Adam optimizer with a binary cross entropy loss function [8 9]. The model was trained using Keras library in python with TensorFlow as a backend [12 13]. Fig. 6 shows the structure of the Deep Neural Network model trained. Equation 1. Represents the cost function used for training.

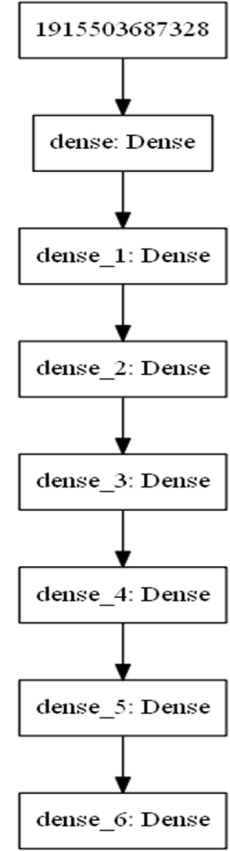

*Fig 6. A visualization for the neural network trained for the given data.*

### D. Cost Function

$$J(\theta) = \frac{1}{m}[\sum_{i=1}^{m}\sum_{k=1}^{K} y_k^i \log h_\theta(x^{(i)}) + (1 - y_i^k)$$

$$\log(1 - (h_\theta(x))_k] + \frac{\lambda}{2m}\sum_{l=1}^{L-1}\sum_{i=1}^{s_l}\sum_{j=1}^{i_l+1}(\theta_{ji}^{(l)})^2$$

Where,

$J(\theta)$ represents the cost function for the given Neural Network.

$\lambda$ represents the regularization parameter.

$\theta_i^{(l)}$ represents the weights of the edges connecting the neurons in the neural network.

### E. Rectified Linear Units

Rectified Linear Unit (ReLU) has an output 0 if the input is negative, and raw output if the input given to the activation function is positive [7]. The ReLU activation can be represented in terms of a mathematic equation as:

$$f(x) = \max(x, 0)$$

Where, $x$ is the input for the activation function $f(x)$. The graphical representation for the ReLU function can be found in Fig 7.

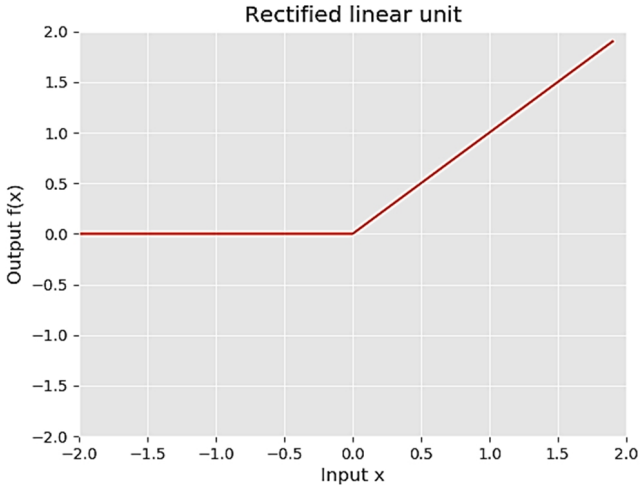

*Fig 7. Graphical representation for the Rectified Linear Unit activation function.*

*F. SoftMax Activation Function*

SoftMax activation function takes a vector of inputs and outputs the categorical probability distribution of the given input. The SoftMax activation makes sure that the outputs are between 0 and 1 like the sigmoid activation function but also divides the outputs in such a way that the sum of the outputs is always equal to 1. The SoftMax function can be mathematically represented as:

$$\sigma(z)_j = \frac{e^{z_j}}{\sum_{k=1}^{K} e^{z_k}}$$

Where,

z represents the given vector as an input.

j indexes the inputs in z as 1,2, 3……, K.

*G. Model Results*

The model so generated provides an accuracy of 95.87% on the training set and 98.30% on the testing set. The confusion matrix for the same can be found in Fig. 7.

The specifications of the machine used to make these predictions are:

- Intel(R) Core (TM) i5-8250U CPU @ 3.40GHz
- CPU(s): 4
- Thread(s) per core: 8
- Core(s) per socket: 2
- Socket(s): 1
- Memory: 8GB, DDR4
- GPU: Nvidia GTX 1050 @ 4GB GDDR5

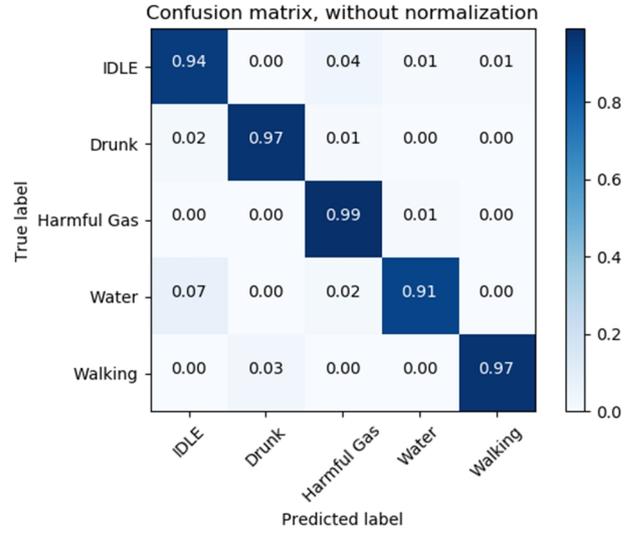

*Fig 8. Confusion matrix of the predictions made by the network.*

*H. Integration of the model into our backend system*

The Model so generated was saved as a file which contained the weights of the layers as an .h5 file. This .h5 file so generated was converted into a TFLite model using TensorFlow TOCO convertor. The TFLite model was uploaded onto the ML Toolkit of the firebase service. This model is then used for the classification of data as received from the sensors in order to predict the type of activity that is being done by then user. The Firebase service makes sure that the model is cached locally on the mobile devices as well in order to make the predictions on the data and display it to the user to avoid any delay in the service and hence reduces the latency.

VIII. CONCLUSION & FUTURE SCOPE

The smart bag is easy to use robust system which will cater the security and health needs of the user. It is simple and easy to use system which is easy to integrated with and android app which tells you about all sensor data, the current body .The confusion matrix figure proves that the accuracy of our neural network is way over 90 percent in each class and the overall test accuracy is 98 percent meaning with overall error rate of only 2 percent the network would give great results in most real time scenario of the bag. Fig, 8,9,10,11 represent the different views of the final system.

In Future we can add several functionalities to the bag. We can add assistance for the disabled using computer vision techniques to detect common objects and notify the person about the same. Solar materials can be used in order to power up the system which is currently being done using an rechargeable battery. Further based on the requirements certain features can be added to the bag for user's ease of use like voice based commands, touch interface on bag, mobile notifications, AUX port for listening to music etc. Apart from the hardware specifications we can collect more data from different persons activity patterns as recorded by the bag and improve the dataset and generate better models from the same.

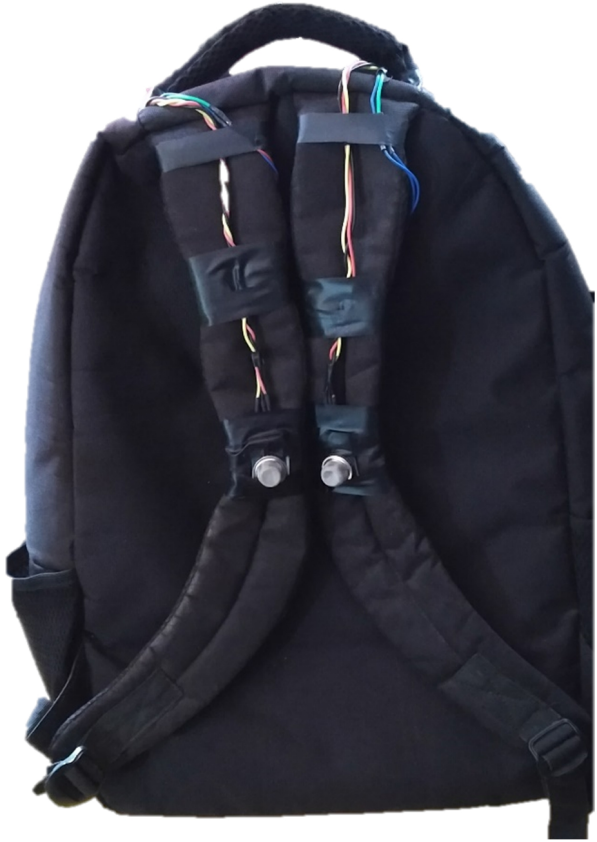

*Fig 9. A rear view of the bag which shows the integration of piezo sensors, SOS button, Gas Sensors on the straps of the bag.*

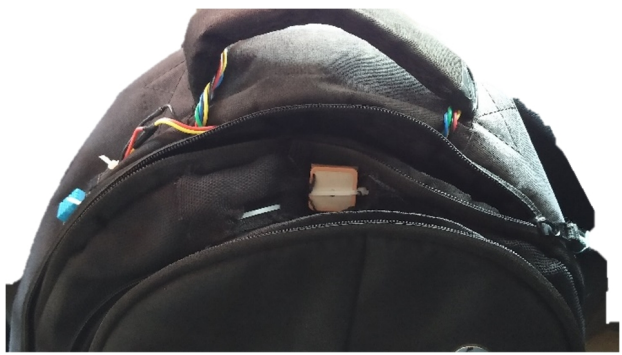

*Fig 10. A top view of the bag showing the integration of the DHT11 temperature and humidity sensor and the GPS Module.*

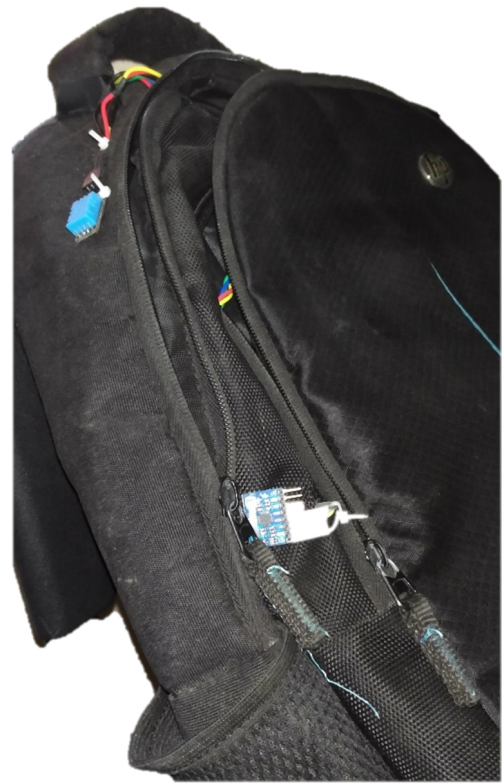

*Fig 11. A side view of the bag showing the integration of the DHT11 temperature and MPU6050 IMU Sensor.*

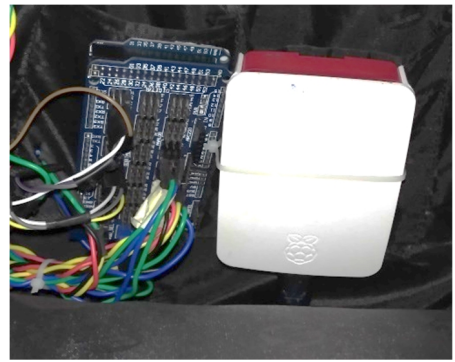

*Fig 12. The view of the front compartment of the bag where the systems and the batteries are integrated for controlling the sensors and acquiring the data from the bag.*